\documentclass[aps,prc,twocolumn,superscriptaddress,nofootinbib,longbibliography,floatfix,10pt]{revtex4-1}

\usepackage{graphicx}
\usepackage{dcolumn}
\usepackage{bm}
\usepackage{morefloats}
\usepackage{multirow}
\usepackage{amssymb}
\usepackage{amsmath}
\usepackage{xcolor} 
\usepackage{longtable}
\usepackage{fix-cm}
\usepackage{mathptmx} 
\usepackage[T1]{fontenc}
\usepackage[colorlinks,allcolors=blue]{hyperref}
\makeatletter
\def\NAT@def@citea{\def\@citea{\NAT@separator}}
\makeatother

\begin{document} 

\title{Quantal diffusion approach for multi-nucleon transfers in Xe + Pb collisions}

\author{S. Ayik}\email{ayik@tntech.edu}
\affiliation{Physics Department, Tennessee Technological University, Cookeville, TN 38505, USA}
\author{B. Yilmaz}
\affiliation{Physics Department, Faculty of Sciences, Ankara University, 06100 Ankara, Turkey}
\author{O. Yilmaz}
\affiliation{Physics Department, Middle East Technical University, 06800 Ankara, Turkey}
\author{A. S. Umar}
\affiliation{Department of Physics and Astronomy, Vanderbilt University, Nashville, TN 37235, USA}

\date{\today}

\begin{abstract} Employing a quantal diffusion description based on the stochastic mean-field (SMF) approach, we analyze the mass distribution of the primary fragments in the collisions of ${}^{136}
\text{Xe}+{}^{208} \text{Pb}$ system at the bombarding energy $E_\text{c.m.} =526$~MeV. This quantal approach provides a good description of the primary fragment distribution without any
adjustable parameter, including the effects of shell structure. 
\end{abstract}


\maketitle

\section{Introduction}
\label{sec1}
It has been recognized that multi-nucleon transfer in heavy-ion collisions provide a suitable mechanism for synthesizing new neutron rich nuclei~\cite{dasso1994,corradi2009,sekizawa2019}.
In particular, multi-nucleon transfer in heavy-ion collisions involving heavy projectile-target combinations
could be utilized for the production of new neutron rich heavy nuclei~\cite{zagrebaev2008c,zagrebaev2011,zagrebaev2012}.
 For this purpose, experimental investigations have been carried out for heavy-ion collisions with heavy projectile-target combinations at near barrier energies~\cite{kozulin2012,watanabe2015,desai2019}. Collisions of heavy systems at near barrier energies predominantly lead to dissipative deep-inelastic reactions and quasi-fission reactions. In dissipative collisions large part of the bombarding energy is converted into internal excitations, and the multi-nucleon transfer occurs between the projectile and target nuclei. Recently, the multi-nucleon transfer mechanism has been investigated for the reaction ${}^{136} \text{Xe}+{}^{208} \text{Pb}$ at bombarding energies $E_\text{c.m.} =423$, 526, and 617~MeV, and di-nuclear mass distributions of the primary fragments have been measured~\cite{kozulin2012}. This system has two unique properties: (i) the neutron shells in the projectile xenon, $N=82$, and the target lead, $N=126$, are closed, and (ii) the $Q_{gg}-$value distributions for nucleon transfers that drive the system toward symmetry and also toward asymmetry have negative values. As a result, the identity of the projectile and target are strongly maintained but data exhibits a broad mass distribution around the projectile and the target masses. 

The multi-dimensional phenomenological Langevin type dynamical approach~\cite{zagrebaev2008c,zagrebaev2011,karpov2017,saiko2019} is quite successful in reproducing many aspects of the data, but the approach is semi-classical and involves a set of adjustable parameters. The mean-field approach of the time-dependent Hartree-Fock (TDHF) theory provides a microscopic approach for describing heavy-ion reaction mechanism at low bombarding energies~\cite{negele1982,simenel2012,simenel2018}. Since several years, the TDHF approach has been used for describing the quasi-fission reactions~\cite{wakhle2014,oberacker2014,umar2015a,umar2016,sekizawa2016,sekizawa2017a}. While the mean-field theory provides a good description for the average values of the collective motion it is not able to accurately describe dynamics of fluctuations for this motion.  The fragment mass distributions provide a good example for the shortcoming of the mean-field description. The TDHF calculations give nearly zero drift for the mass asymmetry, while the dominant aspect of the data is a broad mass distributions around the projectile and target masses resulting from multi-nucleon diffusion mechanism. The description of such large fluctuations requires an approach beyond the mean-field theory.  The time-dependent random phase approximation (TDRPA) approach of Balian and V\'en\'eroni~\cite{balian1984,balian1985,broomfield2008,broomfield2009,simenel2011} provides a possible approach for calculating dispersion of fragment mass distributions. However, this approach has severe technical difficulties in applications to the collisions of asymmetric systems~\cite{williams2018}. Here, we employ the stochastic mean-field (SMF) approach~\cite{ayik2008,lacroix2014} to calculate the mass distribution of the primary fragments in ${}^{136} \text{Xe}+{}^{208} \text{Pb}$ system.  In Sec.~\ref{sec2}, we present a brief description of the quantal nucleon diffusion mechanism based on the SMF approach. In Sec.~\ref{sec3}, we present an analysis of the potential energy surface in the vicinity of the ${}^{136} \text{Xe}+{}^{208} \text{Pb}$ system. The result of calculations for the mass distribution in ${}^{136} \text{Xe}+{}^{208} \text{Pb}$ collisions is reported in Sec.~\ref{sec4}, and conclusions are given in Sec.~\ref{sec5}.

\section{Quantal nucleon diffusion mechanism}
\label{sec2}
\begin{figure}[!hpt]
\includegraphics*[width=4.5cm]{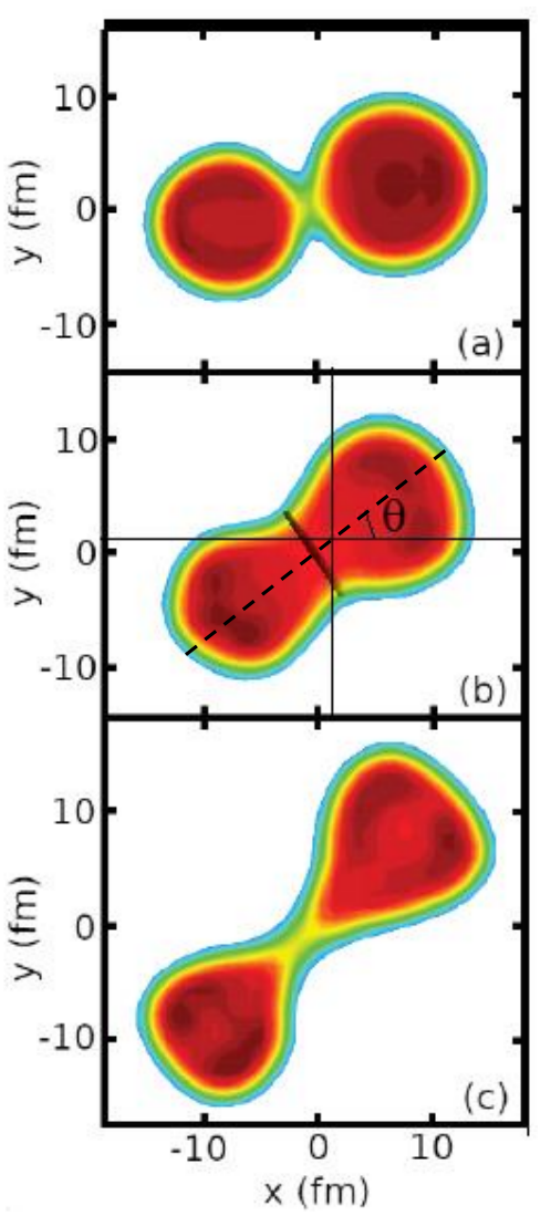}
\caption{(color online) The density profile and the collision geometry of the ${}^{136} \text{Xe}+{}^{208} \text{Pb}$ collisions at $E_\text{c.m.} =526$~MeV with the initial orbital angular momentum $l=100\hbar $ at times (a) $t=300$~fm/c, (b) $t=600$~fm/c, and (c) $t=900$~fm/c.}
\label{fig1}
\end{figure}

As illustrated in previous publications~\cite{ayik2017,ayik2018,yilmaz2018}, the SMF approach gives rise to a quantal Langevin description for the relevant macroscopic variable including the quantal shell effects~\cite{gardiner1991,weiss1999}. Here, we consider nucleon exchange in collisions between heavy nuclei at near barrier energies in which the di-nuclear structure is maintained. We take the neutron $N_{1}^{\lambda}$ and proton $Z_{1}^{\lambda}$ numbers of the projectile-like fragments as the macroscopic variables. In each event $\lambda $, the neutron and proton numbers are determined by integrating the nucleon density over the projectile side of the window between the colliding nuclei,
\begin{align} \label{eq1}
\left(\begin{array}{c} {N_{1}^{\lambda } (t)} \\ {Z_{1}^{\lambda } (t)} \end{array}\right)=\int d^{3}r\;\Theta \left[x'(t)\right]\left(\begin{array}{c} {\rho _{n}^{\lambda } (\vec{r},t)} \\ {\rho _{p}^{\lambda } (\vec{r},t)} \end{array}\right)
\end{align}
where $x'(t)=[y-y_{0} (0)]\sin \theta (t)+\left[x-x_{0} (t)\right]\cos \theta (t)$. The $(x,y)-$plane represents the reaction plane, 
with $x-$axis being the beam direction in the center of mass frame (COM) of the colliding ions. The window plane is perpendicular to the symmetry axis and its orientation is specified by the condition $x'(t)=0$. In this expression, $x_{0} (t)$ and $y_{0} (t)$ denote the coordinates of the window center relative to the origin of the COM frame, $\theta (t)$ is the smaller angle between the orientation of the symmetry axis and the beam direction.  For each impact parameter $b$ or the initial orbital angular momentum, as described in Appendix~A of Ref.~\cite{ayik2018}, by employing the TDHF description, it is possible to determine time evolution of the rotation angle $\theta (t)$ of the symmetry axis.  The coordinates $x_{0} (t)$ and $y_{0} (t)$ of the center point of the window are located at the center of the minimum density slice on the neck between the colliding ions. As an example, Fig.~\ref{fig1} shows the collision geometry in the ${}^{136} \text{Xe}+{}^{208} \text{Pb}$ system at $E_\text{c.m.} =526$ MeV with the initial orbital angular momentum $l=100\hbar $ at times $t=300$~fm/c , $t=600$~fm/c and $t=900$~fm/c. The window plane and symmetry axis of the di-nuclear complex are indicated by thick  and dash lines in frame (b) of this figure. In the following, all quantities are calculated for a given initial orbital angular momentum $l$, but for the purpose of clarity of expressions, we do not attach  the angular momentum label to the quantities. The quantity in Eq.~\eqref{eq1}
\begin{align} \label{eq2}
\rho _{\alpha }^{\lambda } (\vec{r},t)=\sum _{ij\in \alpha }\Phi _{j}^{*\alpha }  (\vec{r},t;\lambda )\rho _{ji}^{\lambda } \Phi _{i}^{\alpha } (\vec{r},t;\lambda )\;,
\end{align}
denotes the neutron and proton number densities in the event $\lambda$ of the ensemble of single-particle density matrices. Here and in the rest of the article, we use the notation $\alpha =n,p$ for the proton and neutron labels.  According to the main postulate of the SMF approach, the elements of the initial density matrix have uncorrelated Gaussian distributions with the zero mean values $\bar{\rho }_{ji}^{\lambda } =\delta _{ji} n_{j} $ and the second moments determined by,
\begin{align} \label{eq3}
\overline{\delta \rho _{ji}^{\lambda } \delta \rho _{i'j'}^{\lambda } }=\frac{1}{2} \delta _{ii'} \delta _{jj'} \left[n_{i} (1-n_{j} )+n_{j} (1-n_{i} )\right]\;,
\end{align}
where $n_{j} $ are the average occupation numbers of the single-particle wave functions of the initial state. At zero initial temperature, the occupation numbers are zero or one, at finite initial temperatures the occupation numbers are given by the Fermi-Dirac functions. Here and below, the bar over the quantity indicates the average over the generated ensemble.

Below, we briefly discuss the derivation of the Langevin equation for neutron and proton numbers of the projectile-like fragments, and for details we refer to Refs.~\cite{ayik2017,ayik2018}. The rate of change of the neutron and the proton numbers for the projectile-like fragment are given by,
\begin{align} \label{eq4}
\frac{d}{dt} \left(\begin{array}{c} {N_{1}^{\lambda } (t)} \\ {Z_{1}^{\lambda } (t)} \end{array}\right)=\int d^{3} r\;\Theta (x')\frac{\partial }{\partial t} \left(\begin{array}{c} {\rho _{n}^{\lambda } (\vec{r},t)} \\ {\rho _{p}^{\lambda } (\vec{r},t)} \end{array}\right)\;.
\end{align}
In this expression, we neglected the terms determined by the rate of change of the position and rotation of the window plane, since tangential and linear velocities of the window are much smaller than the Fermi velocity of nucleons. Using the continuity equation, we obtain a Langevin description for stochastic evolution the neutron and proton numbers of the projectile-like fragments

\begin{align} \label{eq5} \frac{d}{dt} \left(\begin{array}{c} {N_{1}^{\lambda } (t)} \\ {Z_{1}^{\lambda } (t)} \end{array}\right)&=\int d^{3} r\;g(x')\left(\begin{array}{c} {\hat{e}\cdot
\vec{j}_{n}^{\lambda } (\vec{r},t)} \\ {\hat{e}\cdot \vec{j}_{p}^{\lambda } (\vec{r},t)} \end{array}\right)\nonumber\\ &=\left(\begin{array}{c} {v_{n}^{\lambda } (t)} \\ {v_{p}^{\lambda } (t)}
\end{array}\right)\;, \end{align} 
where $\hat{e}$ denotes the unit vector along the symmetry axis with components $\hat{e}_{x} =\cos \theta $ and $\hat{e}_{y} =\sin \theta $.  The quantity
$g(x)=\left(1/\kappa \sqrt{2\pi } \right)\exp \left(-x^{2} /2\kappa ^{2} \right)$ represents a Gaussian with a dispersion $\kappa $, which behaves almost like delta function for sufficiently
small $\kappa $.  In the numerical calculations dispersion of the Gaussian is taken to be on the order of the lattice size $\kappa =1.0$~fm.  The right side of Eq.~\eqref{eq5} defines the fluctuating
drift coefficients $v_{\alpha }^{\lambda } (t)$ for the neutrons and the protons in the event $\lambda $. In the SMF approach the fluctuating current density vector in the event $\lambda$ is given
by, 
\begin{align}\label{eq6} \vec{j}_{\alpha }^{\lambda } (\vec{r},t)=\frac{\hbar }{m} \sum _{ij\in \alpha }Im\left(\Phi _{j}^{*\alpha } (\vec{r},t;\lambda )\vec{\nabla }\Phi _{i}^{\alpha }
(\vec{r},t;\lambda )\rho _{ji}^{\lambda } \right)\;. \end{align}

\subsection{TDHF calculations for mean dynamical path} Equations for the mean values of proton $Z_{1} (t)=\bar{Z}_{1}^{\lambda } (t)$ and neutron $N_{1} (t)=\bar{N}_{1}^{\lambda } (t)$ numbers of
the projectile-like fragments are obtained by taking the ensemble averaging of the Langevin Eq.~\eqref{eq5}.  For small amplitude fluctuations, and using the fact that average values of density
matrix elements are given by the average occupation numbers as $\bar{\rho }_{ji}^{\lambda } =\delta _{ji} n_{j} $, we obtain the usual mean-field results given by the TDHF equations, 
\begin{align}
\label{eq7} \frac{d}{dt} \left(\begin{array}{c} {N_{1} (t)} \\ {Z_{1} (t)} \end{array}\right)=\int d^{3}r\;g(x')\left(\begin{array}{c} {\hat{e}\cdot \vec{j}_{n} (\vec{r},t)} \\ {\hat{e}\cdot
\vec{j}_{p} (\vec{r},t)} \end{array}\right) =\left(\begin{array}{c} {v_{n} (t)} \\ {v_{p} (t)} \end{array}\right). \end{align} 
Here, the mean values of the densities and the currents densities of
neutron and protons are given by, \begin{align} \label{eq8} \rho _{\alpha } (\vec{r},t)=\sum _{h\in \alpha }\Phi _{h}^{*\alpha } (\vec{r},t)\Phi _{h}^{\alpha } (\vec{r},t) \end{align} and
\begin{align} \label{eq9} \vec{j}_{\alpha } (\vec{r},t)=\frac{\hbar }{m} \sum _{h\in \alpha }\text{Im}\left(\Phi _{h}^{*\alpha } (\vec{r},t)\vec{\nabla }\Phi _{h}^{\alpha } (\vec{r},t)\right)\;,
\end{align}
 where the summation $h$ runs over the occupied states originating both from the projectile and the target nuclei. The drift coefficients $v_{p} (t)$ and $v_{n} (t)$ denote the net
proton and neutron currents across the window.

\begin{table}[h]
\caption{Result of TDHF calculations for ${}^{136} \text{Xe}+{}^{208} \text{Pb}$ at $E_\text{c.m.} =526$~MeV for final values of mass and charge of the projectile-like ($A_{1}^{f} $, $Z_{1}^{f} $)
and target-like fragments ($A_{2}^{f} $, $Z_{2}^{f} $), final orbital angular momentum $l_{f}$, total kinetic energy ($TKE$), center of mass $\theta _\text{c.m.} $, laboratory scattering angles
($\theta _{1}^{lab} $,$\theta _{2}^{lab} $), and total excitation energy $E^*$ for a set initial orbital angular momentum $l_{i}$.}
\label{tab1}
\begin{ruledtabular}
\begin{tabular}{|c|c|c|c|c|c|c|c|c|c|c|}
\hline
$l_i\,$($\hbar$) & A$_1^f$ & Z$_1^f$ & A$_2^f$ & Z$_2^f$ & $l_f\,$($\hbar$) & TKE  
& $\theta_{c.m.}$ & E$^*$ & $\theta_1^{lab}$ & $\theta_2^{lab}$ \\
&&&&&&(MeV)&&(MeV)&& \\
\hline
100 & 135 & 53.6 & 209 & 82.4 & 83.1 & 346 & 125  & 185  & 74.8 & 24.2 \\
120 & 135 & 53.9 & 209 & 82.1 & 101  & 349 & 116  & 181  & 67.8 & 28.5 \\
140 & 137 & 54.4 & 207 & 81.6 & 119  & 350 & 116  & 179  & 61.6 & 32.5 \\
160 & 138 & 55.2 & 206 & 80.8 & 131  & 353 & 97.9 & 176  & 55.8 & 36.5 \\
180 & 139 & 55.5 & 205 & 80.5 & 146  & 355 & 90.5 & 172  & 55.7 & 36.6 \\
200 & 137 & 54.9 & 207 & 81.1 & 166  & 348 & 81.3 & 179  & 47.5 & 41.6 \\
220 & 137 & 54.8 & 207 & 81.2 & 177  & 350 & 80.8 & 176  & 45.6 & 43.0 \\
240 & 138 & 55.6 & 206 & 80.4 & 192  & 367 & 79.5 & 160  & 45.2 & 44.6 \\
260 & 137 & 55.1 & 207 & 80.9 & 213  & 397 & 79.1 & 128  & 46.1 & 45.8 \\
280 & 136 & 54.8 & 208 & 81.2 & 238  & 429 & 78.4 & 103  & 46.6 & 47.3 \\
300 & 137 & 54.6 & 207 & 81.4 & 277  & 472 & 77.9 & 53.7 & 47.3 & 49.2 \\
\end{tabular}
\end{ruledtabular}
\end{table}
We carry out TDHF calculations for ${}^{136} \text{Xe}+{}^{208} \text{Pb}$ at $E_\text{c.m.} =526$~MeV for a set initial orbital angular momenta in the range $l=(100-300)\hbar $. This range
of the orbital angular momenta correspond to the data collection range in the laboratory frame ~\cite{kozulin2012}. Table~\ref{tab1} shows the result of TDHF calculations for the final values of mass and charge of the
projectile-like and target-like fragments, the final orbital angular momenta, the total kinetic energy, the center of mass and laboratory scattering angles, and the total excitation energy for a
set initial orbital angular momenta. These calculations and calculations presented in the rest of the paper are performed using the TDHF program developed by Umar et
al.~\cite{umar2006c}. A large part of the initial kinetic energy is dissipated during the collisions. The calculations give very small amount of mass drift, on the order of one mass
unit of neutron and proton drifts at all impact parameters, which is consistent with data. As a result of the neutron shell closures in both projectile and target with $N_{0} =82$ and $N_{0} =126$,
and the due to $Q_{gg}-$ values, the ${}^{136} \text{Xe}+{}^{208} \text{Pb}$ di-nuclear system occupies a local potential minimum state in the (N-Z) plane.  The system has a unique
aspect of strongly preserving its initial di-nuclear structure on the average, but data exhibits remarkably broad mass distribution of the primary fragments.

\subsection{Quantal Langevin equation for neutron and proton diffusion}
Equation~\eqref{eq5} provides a Langevin description for the stochastic evolution the neutron and the proton numbers of the projectile-like fragments. For relatively small fluctuations, we linearize the Langevin equation around the mean evolution.  The drift coefficients $v_{\alpha }^{\lambda } (t)$ fluctuate from event to event due to stochastic elements of the initial density matrix $\delta \rho _{ji}^{\lambda } $ and due to the different sets of the wave functions in different events.  We can represent the fluctuations due to state dependence of the drift coefficients in terms of the fluctuating neutron and proton numbers as $v_{\alpha } (N_{1}^{\lambda } ,Z_{1}^{\lambda } )$. As a result, we can express the linearized Langevin equation as,
\begin{align} \label{eq10}
\frac{d}{dt} \left(\begin{array}{c} {\delta Z_{1} (t)} \\ {\delta N_{1} (t)} \end{array}\right)&=\left(\begin{array}{c} {\frac{\partial v_{p} }{\partial Z_{1} } \left(Z_{1}^{\lambda } -Z_{1} \right)+\frac{\partial v_{p} }{\partial N_{1} } \left(N_{1}^{\lambda } -N_{1} \right)} \\ {\frac{\partial v_{n} }{\partial Z_{1} } \left(Z_{1}^{\lambda } -Z_{1} \right)+\frac{\partial v_{n} }{\partial N_{1} } \left(N_{1}^{\lambda } -N_{1} \right)} \end{array}\right)\nonumber\\
&\quad+\left(\begin{array}{c} {\delta v_{p}^{\lambda } (t)} \\ {\delta v_{n}^{\lambda } (t)} \end{array}\right)\;,
\end{align}
where the derivatives of drift coefficients are evaluated at the mean values $N_{1} $ and $Z_{1} $. The linear limit provides a good approximation for small amplitude fluctuations and it becomes even better if the fluctuations are nearly harmonic around the mean values. The stochastic part $\delta v_{\alpha }^{\lambda } (t)$ of drift coefficients are given by,
\begin{align} \label{eq11}
\delta v_{\alpha }^{\lambda } (t)=\frac{\hbar }{m} \sum _{ij\in \alpha }&\int d^{3} rg(x')\nonumber\\
&\,\times\text{Im}\left(\Phi _{j}^{*\alpha } (\vec{r},t)\stackrel{\frown}{e}\cdot \vec{\nabla }\Phi _{i}^{\alpha } (\vec{r},t)\delta \rho _{ji}^{\lambda } \right)\;.
\end{align}
The variances and the co-variance of neutron and proton distribution are defined as $\sigma _{NN}^{2} (t)=\overline{\left(N_{1}^{\lambda } -N_{1} \right)^{2} }$, $\sigma _{ZZ}^{2} (t)=\overline{\left(Z_{1}^{\lambda } -Z_{1} \right)^{2} }$, and $\sigma _{NZ}^{2} (t)=\overline{\left(N_{1}^{\lambda } -N_{1} \right)\left(Z_{1}^{\lambda } -Z_{1} \right)}$. Multiplying both side of Langevin Eqs.~\eqref{eq10} by $N_{1}^{\lambda } -N_{1} $ and $Z_{1}^{\lambda } -Z_{1} $, and taking the ensemble average, we find evolution of the co-variances are specified by the following set of coupled differential equations~\cite{schroder1981,merchant1982},
\begin{align} \label{eq12}
\frac{\partial }{\partial t} \sigma _{NN}^{2} =2\frac{\partial v_{n} }{\partial N_{1} } \sigma _{NN}^{2} +2\frac{\partial v_{n} }{\partial Z_{1} } \sigma _{NZ}^{2} +2D_{NN}
\end{align}
\begin{align} \label{eq13}
\frac{\partial }{\partial t} \sigma _{ZZ}^{2} =2\frac{\partial v_{p} }{\partial Z_{1} } \sigma _{ZZ}^{2} +2\frac{\partial v_{p} }{\partial N_{1} } \sigma _{NZ}^{2} +2D_{ZZ}
\end{align}
\begin{align} \label{eq14}
\frac{\partial }{\partial t} \sigma _{NZ}^{2} =\frac{\partial v_{p} }{\partial N_{1} } \sigma _{NN}^{2} +\frac{\partial v_{n} }{\partial Z_{1} } \sigma _{ZZ}^{2} +\sigma _{NZ}^{2} \left(\frac{\partial v_{p} }{\partial Z_{1} } +\frac{\partial v_{n} }{\partial N_{1} } \right) .
\end{align}
In these expressions $D_{NN} $ and $D_{ZZ} $ denote the neutron and proton quantal diffusion coefficients which are discussed below. It is well known that the Langevin equation~\eqref{eq10} is equivalent to the Fokker-Planck equation for the correlated distribution function $P(N,Z)$ of the neutron and proton numbers of projectile-like or target-like fragments~\cite{risken1996}. Here, we consider the mass number distribution of the projectile-like and target-like primary fragments. Analytic solution of the Langevin equation for the projectile-like fragments is given by a Gaussian function $P(A,t)$
\begin{align} \label{eq15}
P(A)=\frac{1}{\sqrt{2\pi } \sigma _{AA} } \exp \left[-\frac{1}{2} \left(\frac{A-A_{1} }{\sigma _{AA} } \right)^{2} \right]\;,
\end{align}
where $A_{1} =N_{1} +Z_{1}$ is the mean value of the mass number of the projectile-like fragments and the variance is given by $\sigma _{AA}^{2} =\sigma _{NN}^{2} +\sigma _{ZZ}^{2} +2\sigma _{NZ}^{2} $ . Distribution function of the target-like fragments is given by a similar expression. We should note that the single Gaussian solution for Fokker-Planck equation and hence the Langevin equation is valid when the derivatives of drift coefficients are continuous as approached from left and right of the mean neutron and proton numbers. If the derivative of drift coefficients are discontinuous, which is the case in the ${}^{136} \text{Xe}+{}^{208} \text{Pb}$ system, the mass dispersion in the asymmetric direction $\sigma _{AA}^{<} $ and the symmetric direction $\sigma _{AA}^{>} $ have different values, therefore we cannot represent  the solution of the Langevin Eq.~\eqref{eq10} by a single Gaussian distribution.  In this case, as it is discussed in Sec.~\ref{sec3}, it is possible to represent the solutions of the Langevin equation as a suitable combination of  Gaussian distributions toward asymmetry $P^{<} (A)$ and toward symmetry $P^{>} (A)$.

\subsection{Neutron and proton diffusion coefficients}
\begin{figure}[!hpt]
\includegraphics*[width=8.0cm]{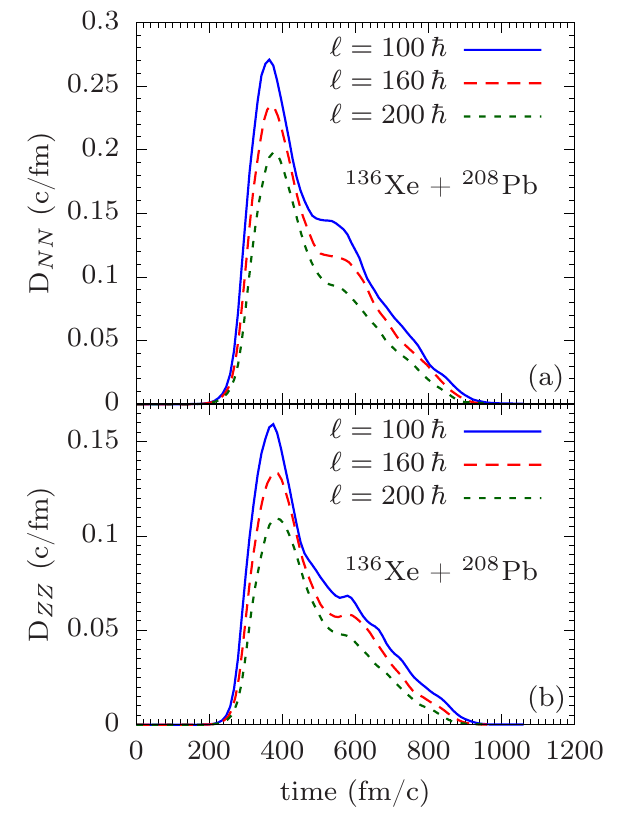}
\caption{(color online) The neutron (a) and proton (b) diffusion coefficients as a function of time for the ${}^{136} \text{Xe}+{}^{208} \text{Pb}$ collisions at $E_\text{c.m.} =526$~MeV with the initial orbital angular momenta $l=100\hbar $, $l=160\hbar $, and $l=200\hbar$.}
\label{fig2}
\end{figure}

The quantal expression of the diffusion coefficients for neutron and proton transfers are determined by the auto-correlation functions of the stochastic part of the drift coefficients as~\cite{ayik2017,ayik2018,yilmaz2018},
\begin{align} \label{eq16}
\int _{0}^{t}dt'\overline{\delta v_{\alpha }^{\lambda } (t)\delta v_{\alpha }^{\lambda } (t')} =D_{\alpha \alpha } (t)\;.
\end{align}
We refer the reader to Refs.~\cite{ayik2017,ayik2018} in which  a detail evaluation of the autocorrelation functions are presented. Here, for completeness of the presentation, we give the results. The quantal expressions of the proton and neutron diffusion coefficients take the form,
\begin{align} \label{eq17}
D_{\alpha \alpha } (t)=&\int _{0}^{t}d\tau  \int d^{3} r \tilde{g}(x')\left(G_{T} (\tau )J_{\bot ,\alpha }^{T} (\vec{r},t-\tau /2)\frac{}{}\right.\nonumber\\
&\qquad\qquad\qquad\quad\left.\frac{}{} +G_{P} (\tau )J_{\bot ,\alpha }^{P} (\vec{r},t-\tau /2)\right)\nonumber\\
&-\int _{0}^{t}d\tau  \text{Re}\left(\sum _{h'\in P,h\in T}A_{h'h}^{\alpha } (t)A_{h'h}^{*\alpha } (t-\tau )\right.\nonumber\\
&\left.\qquad\qquad\quad+\sum _{h'\in T,h\in P}A_{h'h}^{\alpha } (t)A_{h'h}^{*\alpha } (t-\tau ) \right)\;,
\end{align}
where $J_{\bot ,\alpha }^{T} (\vec{r},t-\tau /2)$ represents the sum of the magnitude of current densities perpendicular to the window due to the hole wave functions originating from target,
\begin{align} \label{eq18}
J_{\bot ,\alpha }^{T} (\vec{r},t-\tau /2)=&\frac{\hbar }{m} \sum _{h\in T}\left|\text{Im}\Phi _{h}^{*\alpha } (\vec{r},t-\tau /2)\right.\nonumber\\
&\qquad\times\left(\hat{e}\cdot \vec{\nabla }\Phi _{h}^{\alpha } (\vec{r},t-\tau /2)\right)| \;,
\end{align}
and $J_{\bot ,\alpha }^{P} (\vec{r},t-\tau /2)$  is given by a similar expression in terms of the hole wave functions originating from the projectile. We observe that there is a close analogy between the quantal expression and the classical diffusion coefficient for the random walk problem~\cite{gardiner1991,weiss1999}. The first line in the quantal expression gives the sum of the nucleon currents across the window from the target-like fragment to the projectile-like fragment and from the projectile-like fragment to the target-like fragment, which is integrated over the memory. This is analogous to the random walk problem, in which the diffusion coefficient is given by the sum of the rate for the forward and backward steps.  The second line in the quantal diffusion expression stands for the Pauli blocking effects in nucleon transfer mechanism, which does not have a classical counterpart. The quantities in the Pauli blocking factors are determined by
\begin{align} \label{eq19}
A_{h'h}^{\alpha } (t)=\frac{\hbar }{2m} \int d^{3} r g(x')&\left(\Phi _{h'}^{*\alpha } (\vec{r},t)\hat{e}\cdot \vec{\nabla }\Phi _{h}^{\alpha } (\vec{r},t)\right.\nonumber\\
&\;\left.-\Phi _{h}^{\alpha } (\vec{r},t)\hat{e}\cdot \vec{\nabla }\Phi _{h'}^{*\alpha } (\vec{r},t)\right)\;.
\end{align}
The memory kernel $G_{T} (\tau )$ in Eq.~\eqref{eq19} is given by
\begin{align} \label{eq20}
G_{T} (\tau )=\frac{1}{\sqrt{4\pi } } \frac{1}{\tau _{T} } \exp [-(\tau /2\tau _{T} )^{2} ]\;,
\end{align}
with the memory time determined by the average flow velocity $u_{T} $ of the target nucleons across the window according to $\tau _{T} =\kappa /|u_{T} (t)|$, and $G_{P} (\tau )$ is given by a similar expression. In a previous work, we estimated the memory time to be about $\tau _{T} =\tau _{P} \approx 25$~fm/c, which is much shorter than the contact time of about $600$~fm/c~\cite{ayik2018}. As a result the memory effect is not important for the nucleon diffusion mechanism. We note that the quantal diffusion coefficients are entirely determined in terms of the occupied single-particle wave functions obtained from the TDHF solutions.  According to the non-equilibrium fluctuation-dissipation theorem, the fluctuation properties of the relevant macroscopic variables must be related to the mean properties. Consequently, evaluations of diffusion coefficients in terms of mean-field properties is not surprising. As examples, Fig.~\ref{fig2} shows neutron (a) and proton (b) diffusion coefficients for the ${}^{136} \text{Xe}+{}^{208} \text{Pb}$ system at $E_\text{c.m.} =526$~MeV for the initial orbital angular momenta $l=100\hbar $, $l=160\hbar $, and $l=200\hbar $, as function of time.

\section{Potential energy surface in N-Z plane}
\label{sec3}
\begin{figure}[!hpt]
\includegraphics*[width=7.5cm]{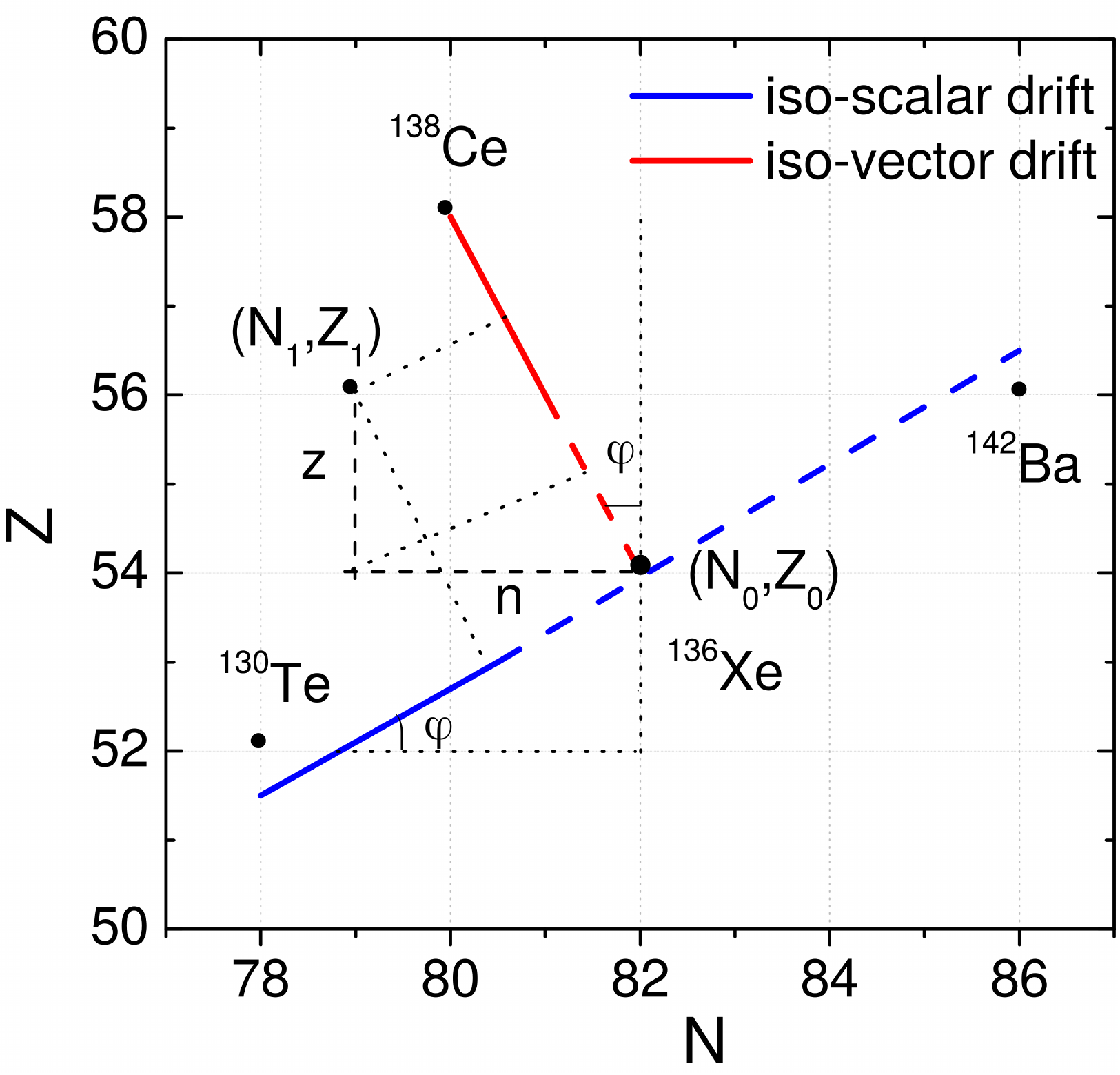}
\caption{(color online) Iso-scalar (thick solid and dash blue line) and iso-vector drift (thick solid and dash red line) paths in the $(N,Z)$- plane.  The locations of ${}^{136} \text{Xe}$, ${}^{138} \text{Ce}$, ${}^{130} \text{Te}$ and ${}^{142} \text{Ba}$ are indicated by black dots.}
\label{fig3}
\end{figure}

For solutions of the co-variances from Eqs.~(\ref{eq12}-\ref{eq14}), in addition to the diffusion coefficients $D_{NN} $ and $D_{ZZ} $, we need to know the rate of change of the drift coefficients. In the ${}^{136} \text{Xe}+{}^{208} \text{Pb}$ system both projectile and target have neutron close shells with $N=82$ and $N=126$, respectively. Furthermore, this di-nuclear system  is placed at the bottom of a local minimum in the potential energy surface. This is evident from the negative $Q_{gg}-$value distribution of the di-nuclear systems in vicinity of ${}^{136} \text{Xe}+{}^{208} \text{Pb}$ . As seen from Fig.~1 of Ref.~\cite{kozulin2012}, $Q_{gg} $-values take increasingly negative values for drifts toward asymmetry, and smaller negative values for drifts toward symmetry. As a result, the system does not exhibit any visible drift between the mass numbers of projectile and target nuclei, but the potential energy surface in $(N,Z)$ plane has a strong influence on the nucleon diffusion mechanism. We consider the projectile-like fragments and indicate  the position of local equilibrium by the neutron and proton numbers of the projectile, $N_{0} =82$ and $Z_{0} =54$ on the $(N,Z)$ plane in Fig.~\ref{fig3}. The charge asymmetry of the projectile ${}^{136} \text{Xe}$ is $(78-52)/136=0.206$. The set of nuclei which have the approximately the same charge asymmetry values are represented by  thick dash-line following  thick solid line (blue line in color) in Fig.~\ref{fig3}. We refer to this line as the iso-scalar drift path. The angle between iso-scalar path and neutron axis is about $\phi =30^{\circ} $, which indicates the iso-scalar line is extending nearly alone the beta stability valley in vicinity of the projectile-like fragments and similarly in the vicinity of target-like fragments. In Fig.~\ref{fig3}, thick dash-line following  thick solid line (red line in color), which is perpendicular to the iso-scalar path, is referred as the iso-vector drift path. We represent the potential energy surface in $\left(N,Z\right)-$plane in terms of two parabolic forms in the iso-scalar and in the iso-vector directions centered at the local equilibrium position of projectile ${}^{136} \text{Xe}$ as,
\begin{align} \label{eq21}
U(N_{1} ,Z_{1} )=&\frac{1}{2} b\left(n\cos \phi -z\sin \phi \right)^{2}\nonumber\\
&+\frac{1}{2} a\left(n\sin \phi +z\cos \phi \right)^{2}\;.
\end{align}
Here $n=N_{0} -N_{1} $ and $z=Z_{1} -Z_{0} $, and $(N_{1} ,Z_{1} )$ indicate neutron and proton number of  a projectile-like fragment in  vicinity of $(N_{0} ,Z_{0} )$. As seen from Fig.~1 in Ref.~\cite{kozulin2012}, $Q_{gg}-$values for nucleon exchanges has a asymmetric distribution, become increasingly negative toward asymmetry, and take smaller negative values toward symmetry. As a result, the parabolic shape of the potential energy, particularly in the iso-scalar direction, can not have a symmetric form. Only for the purpose of clarity we represent the potential parabolas in the symmetric form in Eq.~\eqref{eq21}. However, in particular in the iso-scalar direction, the parabolic form of the potential energy must have an asymmetric shape. It must have a larger curvature in the asymmetry direction than the curvature in the symmetry direction. The situation is analogous to the elastic potential energy of an asymmetric spring.

In order to determine the curvature parameters, we employ the Einstein's relation, which provides a relation between the diffusion and drift coefficients in the transport mechanism of the relevant collective variables and it is often used in the phenomenological approaches~\cite{gardiner1991,weiss1999,schroder1981,merchant1982}.  According to the Einstein's relation, the connection between the neutron $v_{n} (t)$ and proton $v_{p} (t)$ drift coefficients and the corresponding diffusion coefficients are given by,

\begin{align}\label{eq22}
v_{n} (t)&=-\frac{D_{NN} }{T} \frac{\partial U}{\partial N_{1} }\nonumber\\
&=D_{NN} \left(\beta R_{v} (t)\cos \phi +\alpha R_{s} (t)\sin \phi \right)
\end{align}
and
\begin{align} \label{eq23}
v_{p} (t)&=-\frac{D_{ZZ} }{T} \frac{\partial U}{\partial Z_{1} }\nonumber\\
&=D_{ZZ} \left(\beta R_{v} (t)\sin \phi -\alpha R_{s} (t)\cos \phi \right)\;.
\end{align}
Here, the temperature is absorbed into the curvature parameters $\beta =b/T$,  $\alpha =a/T$, and the quantities $R_{v} (t)=n\cos \phi -z\sin \phi $, $R_{s} (t)=n\sin \phi +z\cos \phi $ represent  the distances  of an arbitrary  fragment $(N_{1} ,Z_{1} )$ located in the vicinity of the  projectile from the iso-vector and the iso-scalar lines, respectively. Because of the analytical form, we can readily calculate the derivatives of the drift coefficients to give,
\begin{align} \label{eq24}
\partial v_{n} (t)/\partial N_{1} &=-D_{NN} \left(\beta \cos ^{2} \phi +\alpha \sin ^{2} \phi \right), \\
\label{eq25}
\partial v_{n} (t)/\partial Z_{1} &=+D_{NN} \left(\alpha -\beta \right)\cos \phi \sin \phi,  \\
\label{eq26}
\partial v_{p} (t)/\partial Z_{1} &=-D_{ZZ} \left(\beta \sin ^{2} \phi +\alpha \cos ^{2} \phi \right), \\
\label{eq27}
\partial v_{p} (t)/\partial N_{1} &=+D_{ZZ} \left(\alpha -\beta \right)\cos \phi \sin \phi\;. 
\end{align}
In principle, it is possible to determine the curvature parameters and hence the derivatives of the drift coefficients from the mean-drift path calculated in the TDHF approach. However, this does not work in the collision of the ${}^{136} \text{Xe}+{}^{208} \text{Pb}$ system. The mean values of neutron and proton numbers of projectile-like fragments $(N_{1} \approx N_{0} ,Z_{1} \approx Z_{0} )$ are nearly equal to their initial values, and similarly for the projectile-like fragments. Therefore, Eqs.~(\ref{eq22},\ref{eq23}) do not allow to determine the curvature parameters from drift information of the system.

\subsection{Curvature parameters for the potential energy parabola}
For determining the curvature parameters, we choose two nearby systems ${}^{130} \text{Te}+{}^{214} \text{Po}$ and ${}^{138} \text{Ce}+{}^{206} \text{Pt}$.  The total mass numbers of both systems equal to the total mass number of ${}^{136} \text{Xe}+{}^{208} \text{Pb}$. The neutron number and proton number of  ${}^{130} \text{Te}$ are smaller than ${}^{136} \text{Xe}$ by four units and two units, respectively. It has a charge asymmetry of $0.200$, which is nearly the same as for the ${}^{136} \text{Xe}$, and as a result it is located very near to the iso-scalar line as indicated in Fig.~\ref{fig3}. On the other hand, in ${}^{138} \text{Ce}$ the neutron number is smaller by two units and proton number is larger by four units than${}^{136} \text{Xe}$, and it is located on the iso-vector line as indicated in Fig.~\ref{fig3}. We carry out the SMF calculations for the ${}^{130} \text{Te}+{}^{214} \text{Po}$ and the ${}^{138} \text{Ce}+{}^{206} \text{Pt}$ at the same $E_\text{c.m.} =526$ MeV and for the initial orbital angular momentum $l=100\hbar $.

\begin{figure}[!hpt]
\includegraphics*[width=8.0cm]{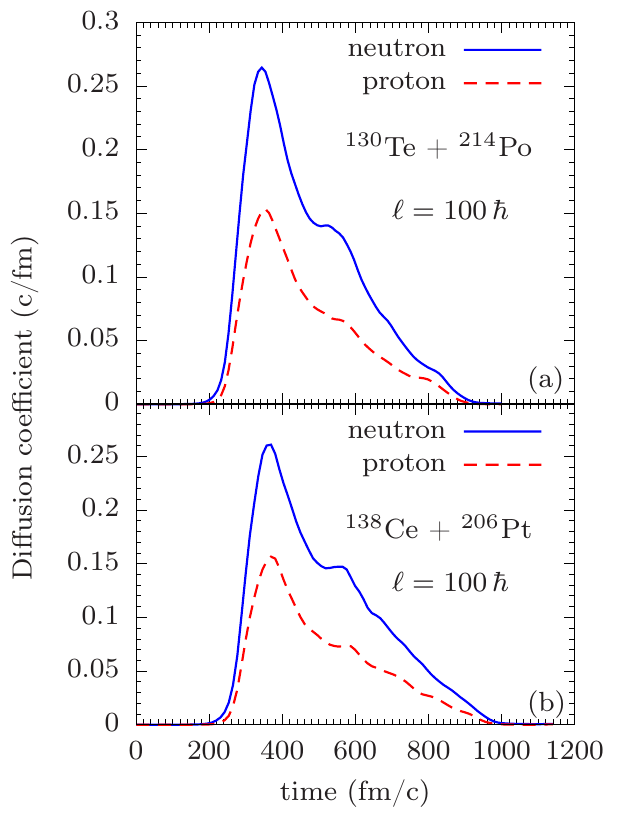}
\caption{(color online) The neutron and proton diffusion coefficients as a function of time for the ${}^{130} \text{Te}+{}^{214} \text{Po}$ (a) and for the ${}^{138} \text{Ce}+{}^{206} \text{Pt}$ (b) collisions at the $E_\text{c.m.} =526$ MeV  with the initial orbital angular momentum $l=100\hbar $.}
\label{fig4}
\end{figure}
\begin{figure}[!hpt]
\includegraphics*[width=8.0cm]{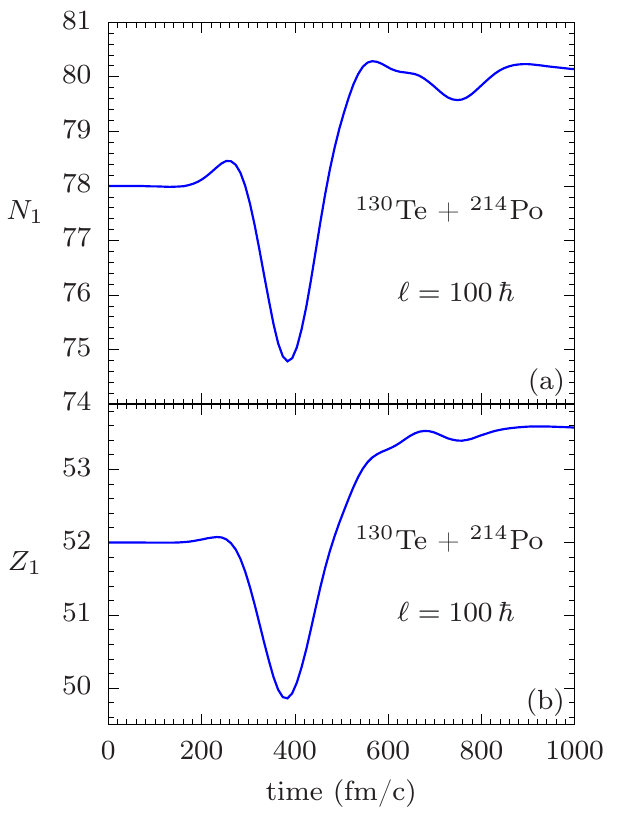}
\caption{(color online) The neutron (a) and proton (b) drift paths of the projectile-like fragments as a function of time for the ${}^{130} \text{Te}+{}^{214} \text{Po}$ collisions with the $E_\text{c.m.} =526$~MeV  at the initial orbital angular momentum $l=100\hbar $.}
\label{fig5}
\end{figure}
Figure~\ref{fig4} shows the neutron and proton diffusion coefficients for the ${}^{130} \text{Te}+{}^{214} \text{Po}$ system (a) and the ${}^{138} \text{Ce}+{}^{206} \text{Pt}$ system (b) at $E_\text{c.m.} =526$~MeV and orbital angular momentum $l=100\hbar $. In these collisions the contact starts at 200~fm/c, fragments separate at around 800~fm/c. In the TDHF description, the potential energy surface involves the full effect of the shell structure and therefore has a complex shape. Figure~\ref{fig5} shows the result of the TDHF calculations of the mean values of the neutron and proton numbers of the projectile-like fragments as a function of time for ${}^{130} \text{Te}+{}^{214} \text{Po}$. The system initially drifts toward asymmetry along the iso-scalar line until about 400~fm/c. As the system heats up the shell effects disappear, the system reverses direction, drifts along the symmetry towards the local minimum and before reaching the minimum location of ${}^{136} \text{Xe}$, it  breaks up. By eliminating $\alpha $ from Eq.~\eqref{eq22} and Eq.~\eqref{eq23}, we can derive an expression for $\beta $ in terms of drift and diffusion coefficients. We use the drift information of ${}^{130} \text{Te}+{}^{214} \text{Po}$ to determine the average value of the iso-scalar curvature parameter toward asymmetry direction $\beta _{<} $ in the time interval from $t_{1} =480$~fm/c to $t_{2} =540$~fm/c. This interval is indicated by thick blue line on the iso-scalar drift path in Fig.~\ref{fig3}, and it approximately corresponds to the average taken over maximum overlap of the colliding nuclei during the iso-scalar drift. In this manner, we estimate the average value of the curvature parameter in the asymmetric side of the of ${}^{136} \text{Xe}$ along the iso-scalar path as,
\begin{align} \label{eq29}
\beta _{<} &=\frac{1}{\Delta t} \int _{t_{1} }^{t_{2} }d\tau \frac{1}{R_{v} (\tau )} \left(\frac{v_{n} (\tau )}{D_{NN} (\tau )} \cos \phi +\frac{v_{p} (\tau )}{D_{ZZ} (\tau )} \sin \phi \right)\nonumber\\
&=0.127\;.
\end{align}
Here, $\Delta t=t_{2} -t_{1} $ and distance $R_{v} $ is evaluated with $N_{1} (t),Z_{1} (t)$ on the iso-scalar path. The drift coefficients are determined from rate of change of the neutron and
proton numbers $v_{n} =\partial N_{1} /\partial t$ and $v_{p} =\partial Z_{1} /\partial t$ in Fig.~\ref{fig5}.

\begin{figure}[!hpt]
\includegraphics*[width=8.0cm]{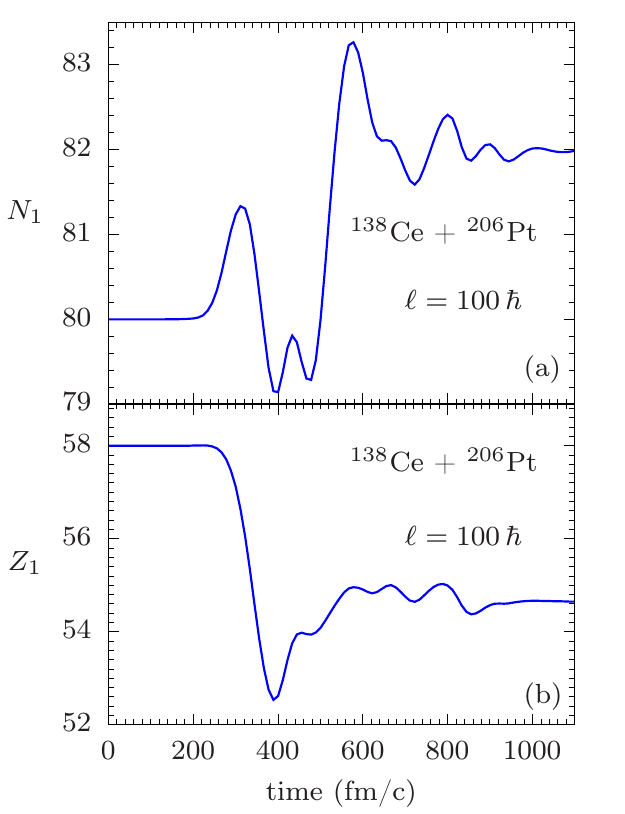}
\caption{(color online) The neutron (a) and proton (b) drift paths of the projectile-like fragments as a function of time in the ${}^{138} \text{Ce}+{}^{206} \text{Pt}$ collisions with the $E_\text{c.m.} =526$~MeV  at the initial orbital angular momentum $l=100\hbar $.}
\label{fig6}
\end{figure}
Figure~\ref{fig6} shows the result of the TDHF calculations of the mean values of the neutron and proton numbers of the projectile-like fragments as a function of time in collision of ${}^{138} \text{Ce}+{}^{206} \text{Pt}$.  In this system time evolution is more complex, but we can recognize a rapid drift during the time interval (200 -- 400)~fm/c along the iso-vector path toward the iso-scalar line and curving toward the asymmetry direction. As the system heats up the shell effects disappear, the system reverses direction, drifts along the symmetry toward the local minimum and almost reaches the minimum location of ${}^{136} \text{Xe}$ before it breaks-up at around 800~fm/c. By eliminating $\beta $ from Eq.~\eqref{eq22} and Eq.~\eqref{eq23}, we can derive an expression for $\alpha $ in terms of drift and diffusion coefficients. We use the drift information of the system ${}^{138} \text{Ce}+{}^{206} \text{Pt}$ to determine the average value of the iso-vector curvature parameter in the time interval from $t_{1} =210$~fm/c to $t_{2} =310$~fm/c.  This time interval approximately correspond to the average taken over the maximum overlap during the drift along the iso-vector path. The projection of this interval on the iso-vector line is indicated by a thick red line in Fig.~\eqref{eq3}. We evaluate the average value of the curvature $\alpha $ over this interval as,
\begin{align} \label{eq30}
\alpha &=\frac{1}{\Delta t} \int _{t_{1} }^{t_{2} }d\tau \frac{1}{R_{s} (\tau )} \left(\frac{v_{n} (\tau )}{D_{NN} (\tau )} \sin \phi -\frac{v_{p} (\tau )}{D_{ZZ} (\tau )} \cos \phi \right) \nonumber\\
&=0.143\;.
\end{align}
Here, $\Delta t=t_{2} -t_{1} $ and distance $R_{s} $ is evaluated with $N_{1} (t)$, $Z_{1} (t)$ on the iso-vector path. The drift coefficients are determined from rate of change of the neutron and proton numbers in Fig.~\ref{fig6}.

Since ${}^{138} \text{Ce}+{}^{206} \text{Pt}$ system rapidly drifts toward the charge symmetry of the iso-scalar path, the asymmetry of the curvature parameter in the iso-vector direction does not have an important effect on the diffusion mechanism. Therefore, we neglect the asymmetry effect in the potential energy in the iso-vector direction. For determining the iso-scalar curvature parameter toward the symmetry direction $\beta_{>} $, we consider the associate system ${}^{142} \text{Ba}+{}^{202} \text{Hg}$.  The charge asymmetry of nuclei ${}^{139} \text{Te}$ and ${}^{142} \text{Ba}$ have nearly the same value  and are located on the iso-scalar path equal distance away from ${}^{136} \text{Xe}$ as indicated in Fig.~\ref{fig3}. Similarly, charge asymmetry of nuclei ${}^{214} \text{Po}$ and ${}^{202} \text{Hg}$ have nearly the same value  and are located close to the iso-scalar path equal distance away from ${}^{208} \text{Pb}$.  In order to save computing time, rather than carrying out the SMF calculations for ${}^{142} \text{Ba}+{}^{202} \text{Hg}$ system, we estimate the iso-scalar curvature parameter toward the symmetry direction $\beta_{>} $,  with the help of the $Q_{gg} -$value distribution along the iso-scalar path. The $Q_{gg} -$value of  ${}^{130} \text{Te}+{}^{214} \text{Po}$ system relative to ${}^{136} \text{Xe}+{}^{208} \text{Pb}$ is $Q_{gg} =-16.2$~MeV.  The associate system ${}^{142} \text{Ba}+{}^{202} \text{Hg}$  has a $Q_{gg} =-2.99$~MeV value relative to ${}^{136} \text{Xe}+{}^{208} \text{Pb}$.  We estimate the potential energy and therefore curvature parameter along symmetry direction with the ratio of the $Q_{gg} -$ values to give $\beta_{>} =\beta_{<} (2.99/16.2)=0.023$. For a heavy di-nuclear system the rotational kinetic energy depends on the mass asymmetry variable in a smooth manner~\cite{merchant1982}. As a result, the effect of the rotational energy on the curvature parameters is very small, and hence in the orbital angular momentum range $l=\left(100-300\right)\hbar $ of the di-nuclear system ${}^{136} \text{Xe}+{}^{208} \text{Pb}$, the average value of the curvature parameters of the parabolic forms have approximately same magnitudes during the maximum overlap of the colliding nuclei. Furthermore, the curvature parameters should be proportional to the window area of the colliding nuclei for each orbital angular momentum. In order to take into account this window effect, we multiply the average values of the curvature parameters with a form factor $\beta_{<}^{l} (t)=\beta_{<}\, F_{l} (t)$, $\beta_{>}^{l} (t)=\beta_{>}\, F_{l} (t)$ and $\alpha _{l} (t)=\alpha F_{l} (t)$. We take this form factor to be the ratio of neutron diffusion coefficients of the ${}^{136} \text{Xe}+{}^{208} \text{Pb}$ for each  orbital angular momentum,
\begin{align} \label{eq31}
F_{l} (t)=D{}_{l} (t)/D{}_{l} (t_{m})\;.
\end{align}
The ratio of the neutron diffusion coefficients provide a measure for the  ratio of the window area at time $t$ to the maximum window area at time $t_{m} $.

\section{Fragment mass distribution in ${}^{136} \text{Xe}+{}^{208} \text{Pb}$}
\label{sec4}

\begin{figure}[!hpt]
\includegraphics*[width=8.0cm]{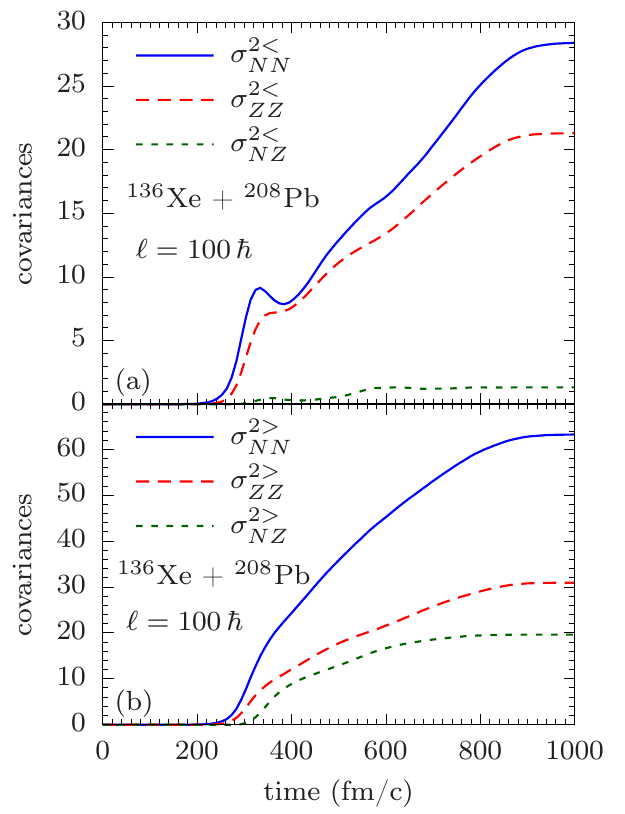}
\caption{(color online) The neutron, proton, and mixed variances as a function of time in the ${}^{136} \text{Xe}+{}^{208} \text{Pb}$ collisions with the $E_\text{c.m.} =526$ MeV  at the initial orbital angular momentum $l=100\hbar $, towards asymmetry (a) and  towards symmetry (b).}
\label{fig7}
\end{figure}

\begin{table}[h]
\caption{Asymptotic values of the variances, the co-variances of neutrons and protons, the mass dispersions in symmetric $\sigma _{AA}^>$ and in asymmetric $\sigma _{AA}^<$ directions, and the dispersion $\bar{\sigma }_{AA} $ of the middle Gauss functions for a set of initial orbital angular momentum in the interval $l=(100-300)\hbar $ in ${}^{136} \text{Xe}+{}^{208} \text{Pb}$  collisions at $E_\text{c.m.} =526$~MeV.}
\label{tab2}
\begin{ruledtabular}
\begin{tabular}{|c|c|c|c|c|c|c|c|c|c|}
\hline
$l_i\,$($\hbar$) & $\sigma_{NN}^{2<}$ & $\sigma_{ZZ}^{2<}$ & $\sigma_{NZ}^{2<}$ & $\sigma_{AA}^<$ & $\sigma_{NN}^{2>}$ & $\sigma_{ZZ}^{2>}$ & $\sigma_{NZ}^{2>}$ & $\sigma_{AA}^>$ & $\overline{\sigma}_{AA}$ \\
\hline 
100 & 28.4 & 24.8 & 1.32 & 7.47 & 63.2 & 30.9 & 19.6 & 11.6 & 9.57 \\
120 & 28.7 & 21.1 & 1.40 & 7.26 & 62.8 & 30.5 & 19.2 & 11.5 & 9.43 \\
140 & 29.2 & 21.2 & 1.47 & 7.30 & 62.5 & 30.1 & 18.7 & 11.4 & 9.41 \\
160 & 29.4 & 21.3 & 1.46 & 7.32 & 61.5 & 29.8 & 17.9 & 11.3 & 9.35 \\
180 & 29.6 & 21.5 & 1.36 & 7.33 & 59.7 & 29.2 & 16.6 & 11.1 & 9.24 \\
200 & 30.4 & 21.8 & 1.29 & 7.40 & 57.7 & 28.4 & 14.8 & 10.8 & 9.12 \\
220 & 30.2 & 21.2 & 1.23 & 7.33 & 53.3 & 26.3 & 12.2 & 10.2 & 8.79 \\
240 & 28.7 & 19.3 & 1.10 & 7.08 & 45.9 & 22.6 & 8.61 & 9.26 & 8.19 \\
260 & 26.4 & 16.1 & 0.79 & 6.64 & 36.5 & 17.7 & 4.74 & 7.98 & 7.37 \\
280 & 21.5 & 11.5 & 0.42 & 5.81 & 25.5 & 11.9 & 1.68 & 6.38 & 6.08 \\
300 & 12.9 & 5.81 & 0.15 & 4.35 & 13.8 & 5.77 & 0.30 & 4.49 & 4.39 \\
\end{tabular}
\end{ruledtabular}
\end{table}
We determine the co-variances of the neutron $\sigma _{NN}^{2} (t)$ and the proton $\sigma _{ZZ}^{2} (t)$ variances and the mixed co-variance $\sigma _{NZ}^{2} (t)$ by solving the differential equations in Eqs.~(\ref{eq12}-\ref{eq14}).  Because of  different curvature parameters in symmetric and asymmetric directions of the iso-scalar path, the variances and the co-variances have different values in symmetric and asymmetric directions. As an example, Fig.~\ref{fig7} shows the variances and the co-variance as a function of time for the initial orbital angular momentum $l=100\hbar $.   We find the dispersion of the mass number distributions  using the expression $\sigma _{AA}^{2} (t)=\sigma _{NN}^{2} (t)+\sigma _{ZZ}^{2} (t)+2\sigma _{NZ}^{2} (t)$.  Table~\ref{tab2} shows the asymptotic values of the variances, the co-variances and the mass dispersions in symmetric $\sigma _{AA}^>$ and in asymmetric $\sigma _{AA}^<$ directions for a set of initial orbital angular momentum in the interval $l=(100-300)\hbar $

The mass number distributions in the asymmetry direction and the symmetry direction are determined by the Gauss functions
\begin{align} \label{eq32}
P_{l}^{<} (A-A_{l} )=\frac{1}{\sqrt{2\pi } } \frac{1}{\sigma _{AA}^{<} (l)} =\exp \left[-\frac{1}{2} \left(\frac{A-A_{l} }{\sigma _{AA}^{<} (l)} \right)^{2} \right]\;,
\end{align}
and
\begin{align} \label{eq33}
P_{l}^{>} (A-A_{l} )=\frac{1}{\sqrt{2\pi } } \frac{1}{\sigma _{AA}^{>} (l)} =\exp \left[-\frac{1}{2} \left(\frac{A-A_{l} }{\sigma _{AA}^{>} (l)} \right)^{2} \right]\;,
\end{align}
where, $A_{l} $ denotes the mean mass number of the projectile-like or the target-like fragments. Because of the asymmetric dispersions, these distribution functions do not match at the mean value of the mass number. In order to provide an approximate analytical description for the solution of the Langevin Eq.~\eqref{eq10} we smoothly combine the left and right Gauss functions. We determine the right and left intersection points $A_{l}^{0} (R)$, $A_{l}^{0} (L)$ by matching the Gauss functions $P_{l}^{<} (A_{l}^{0} -A_{l} )=P_{l}^{>} (A_{l} -A_{l}^{0} )=\bar{P}_{l} $, and smoothly  join the left and right Gauss functions between the intersection points by a middle Gauss function,
\begin{align} \label{eq34}
\bar{P}_{l} (A-A_{l} )=\bar{P}_{l} \exp \left[-\frac{1}{2} \left(\frac{A-A_{l} }{\bar{\sigma }_{AA} (l)} \right)^{2} +\frac{1}{2} \left(\frac{\Delta A_{l} }{\bar{\sigma }_{AA} (l)} \right)^{2} \right]\;,
\end{align}
where $\Delta A_{l} =A_{l} -A_{l}^{0} (L)=A_{l}^{0} (R)-A_{l} $. The dispersions of the middle Gauss functions are determined by requiring  the entire distribution is normalized to one for each orbital angular momentum. This requirement is given by the integral relation,
\begin{align} \label{eq35}
\int _{0}^{\Delta A_{l} }\frac{1}{2}  dA\left[P_{l}^{<} (A)+P_{l}^{>} (A)\right]=\int _{0}^{\Delta A_{l} }dA \bar{P}_{l} (A)\;.
\end{align}
Dispersions $\bar{\sigma }_{AA} (l)$ of the middle Gauss functions determined from this  requirement are listed in the last column of Table~\ref{tab2}.  The dispersion of the middle Gauss functions are approximately equal to the average values of the left and right  dispersions, $\bar{\sigma }_{AA} (l)\approx \left(\sigma _{AA}^{<} (l)+\sigma _{AA}^{>} (l)\right)/2$. As a result, the middle Gauss functions  describe nearly the average values of the left and the right Gauss functions in the intersection intervals, which have about $\Delta A_{l} \approx 8-10$ nucleons range from the mean values.

\begin{figure}[!hpt]
\includegraphics*[width=8.0cm]{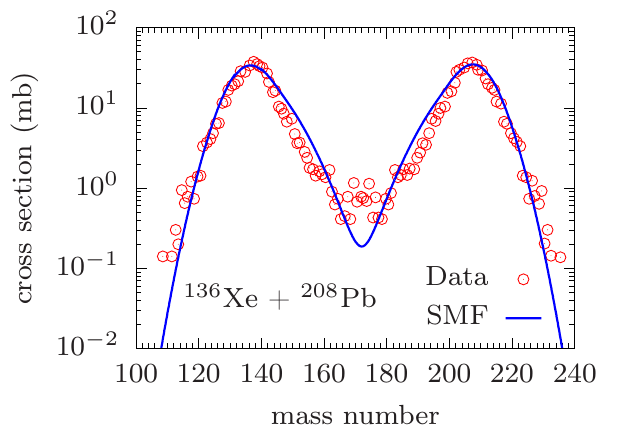}
\caption{(color online) Line shows mass number distribution of the primary fragments in collisions of ${}^{136} \text{Xe}+{}^{208} \text{Pb}$  system at $E_\text{c.m.} =526$~MeV. Data is taken from ~\cite{kozulin2012}.}
\label{fig8}
\end{figure}
We calculate the cross-section for production of  a fragment with the mass number $A$ using the standard expression,
\begin{align} \label{eq36}
\sigma (A)=\frac{\pi \hbar ^{2} }{2\mu E_\text{c.m.} } \sum _{l_{\min } }^{l_{\max } }(2l+1)P_{l} (A)\;,
\end{align}
where $P_{l} (A)=P_{1,l} (A)+P_{2,l} (A)$ is the probability distribution of the total fragments and the summation runs over from the $l_{\min } =100\hbar$ to $l_{\max } =300\hbar$. As shown in Table~\ref{tab1}, the total excitation energy of TDHF calculations at maximum $l_{\max } =300\hbar$ value is $E^{*}=53.7$~MeV, which is nearly the same for the $TKE=40.0$~MeV cut-off in the experimental fragment mass distribution in Fig.~\ref{fig8}.  Here, $P_{1,l} (A)$ and $P_{2,l} (A)$ denote the probability distribution functions of the projectile-like and the target-like fragments for the initial orbital angular momentum $l$.  The distribution functions of the projectile-like fragments and the target-like fragments are determined according to,
\begin{align} \label{eq37}
P_{1,l} (A)=\left\{\begin{array}{ccc} {P_{l}^{<} (A-A_{1l} )} & {A\le A_{1l} -\Delta A_{l} } & {} \\ {\bar{P}_{l} (A-A_{1l} )} & {A_{1l} -\Delta A_{l} \le A\le A_{1l} +\Delta A_{l} } & {} \\ {P_{l}^{>} (A-A_{1l} )} & {A\ge A_{1l} +\Delta A_{l} } & {} \end{array}\right.\;,
\end{align}
and
\begin{align} \label{eq38}
P_{2,l} (A)=\left\{\begin{array}{ccc} {P_{l}^{<} (A-A_{2l} )} & {A\le A_{2l} -\Delta A_{l} } & {} \\ {\bar{P}_{l} (A-A_{2l} )} & {A_{2l} -\Delta A_{l} \le A\le A_{2l} +\Delta A_{l} } & {} \\ {P_{l}^{>} (A-A_{2l} )} & {A\ge A_{2l} +\Delta A_{l} } & {} \end{array}\right.\;.
\end{align}
In these expressions $A_{1l} $ and $A_{2l} $ indicate the mean mass values of the projectile-like and the target-like primary fragments, respectively. We consider these distributions as averages over $20$ angular momentum unit intervals and carry out the summation as follows,
\begin{align} \label{eq39}
\sigma (A)=\frac{\pi \hbar ^{2} }{2\mu E_\text{c.m.} } \sum _{n=0}^{n=10}10\times (2l_{n} +1)P_{l_{n} } (A)\;.
\end{align}
Here, $l_{n} =100+20\times n$  denotes the average orbital angular momentum quantum number in the $20$ unit intervals. Since the total probability $P_{l_{n} } (A)$ is normalized to two, the factor $10$ appear rather than $20$ in  front of $(2l_{n} +1)$.  Fig.~\ref{fig8} presents a comparison of the  calculated cross-section for the primary fragment production indicated by solid line in collision of the ${}^{136} \text{Xe}+{}^{208} \text{Pb}$  system at $E_\text{c.m.} =526$ MeV and  experimental cross-sections of Kozulin et al.~\cite{kozulin2012} are indicated  by circles.

\section{Conclusions}
\label{sec5}
We carry out an investigation of mass distributions of the primary fragments produced in the collisions of the ${}^{136} \text{Xe}+{}^{208} \text{Pb}$  system at $E_\text{c.m.} =526$ MeV. We calculate the mass distribution employing a quantal nucleon diffusion mechanism based on the SMF approach. The diffusion coefficients of neutrons and protons, which describes the fluctuation aspects of the mass distribution, are determined entirely in terms of the occupied single particle states of the TDHF equations and  they do not involve any adjustable parameters other than the standard parameters of the effective Skyrme interaction. The evaluation of the transport coefficients in terms of the mean-field properties is consistent with the fundamental idea of the non-equilibrium fluctuation-dissipation theorem.  The potential energy surface in the $(N,Z)$- plane has an important effect on the diffusion mechanism. As a result of the neutron shell closure of projectile and target  with
$N=82$ and $N=126$, respectively, and due the $Q_{gg} -$value distributions, the ${}^{136} \text{Xe}+{}^{208} \text{Pb}$ system is located at a local potential energy minimum position in the  $(N,Z)-$ plane. We parameterize the potential energy in the vicinity of this local minimum in terms of two parabolic forms along the iso-scalar and iso-vector directions. We determine the curvature parameters of the parabolic forms by carrying out the SMF calculations for two nearby systems ${}^{130} \text{Te}+{}^{214} \text{Po}$ and ${}^{138} \text{Ce}+{}^{206} \text{Pt}$, and utilizing the $Q_{gg} -$value information of the systems ${}^{130} \text{Te}+{}^{214} \text{Po}$ and ${}^{142}\text{Ba}+{}^{202} \text{Hg}$, which are located symmetrically on the $(N,Z)-$ plane along the iso-scalar direction.  As seen in Fig.~\ref{fig8}, the quantal diffusion calculations based on the SMF approach provides a very good description of the mass distribution of the primary fragments of data reported by the Kozulin et al.~\cite{kozulin2012} without any adjustable parameters.

\begin{acknowledgments}
S.A. gratefully acknowledges the IPN-Orsay and the Middle East Technical University for warm hospitality extended to him during his visits. S.A. also gratefully acknowledges useful discussions with D. Lacroix, and very much thankful to his wife F. Ayik for continuous support and encouragement. This work is supported in part by US DOE Grant Nos. DE-SC0015513 and DE-SC0013847, and in part TUBITAK Grant No. 117F109.
\end{acknowledgments}

\bibliography{VU_bibtex_master}

\end{document}